\documentclass[aps,physrev, preprint,preprintnumbers,longbibliography,a4paper,12pt,titlepage, nofootinbib,superscriptaddress]{revtex4-2}
\usepackage[utf8]{inputenc}
\usepackage[version=3]{mhchem}
\usepackage{braket}
\usepackage{amsmath,amssymb}
\usepackage{MnSymbol}
\usepackage{mathtools}
\usepackage{physics}
\usepackage{tensor}
\usepackage{empheq}
\usepackage{eufrak}
\usepackage{mathrsfs}
\usepackage{collectbox}
\usepackage{csquotes}
\usepackage{graphicx}
\usepackage{dcolumn}
\usepackage{bm}
\usepackage[colorlinks,breaklinks=true]{hyperref}
\hypersetup{
     colorlinks=true,
     linkcolor=blue,
     filecolor=blue,
     citecolor = black,      
     urlcolor=blue,
     }

\begin{document}

\preprint{}

\title{Quantum corrections to the Lorentz algebra due to mixed gravitational-$U(1)$-chiral anomalies.} 
\author{Sandeep S. Cranganore}
\email{sandeep.cranganore@tuwien.ac.at}
\affiliation{Institute for Theoretical Physics, University of Cologne, 50923 Cologne, Germany}
\affiliation{TU Wien - Atominstitut,
Stadionallee 2, 1020 Vienna, Austria}
\date{\today}
\begin{abstract}
We calculate the quantum corrections to the Lorentz algebra for chiral Weyl fermions interacting with an external $U(1)$ gauge field in a background Riemann-Cartan (RC) spacetime. This was achieved by setting up the equal-time commutation relations (ETCR) for the canonical spin current of chiral Weyl fermions. Furthermore, these quantum corrections lead to an order sensitive commutator, i.e., swapping the Lorentz generators in the commutator doesn't merely lead to a sign change, but rather a completely different correction term to the Lorentz algebra. Thus, the algebra of Lorentz is altered due to anomalies associated with chiral particles.
\end{abstract}

\maketitle


\section*{NOTATIONS}
Spacetime coordinates will be labeled with Latin indices $i, j, k...=0, 1, 2, 3$. Spatial coordinates will be denoted by a, b,....= 1, 2, 3. The frame (tetrad) fields are denoted as $e_\alpha$ with components $\tensor{e}{^i_\alpha}$, where $\alpha$, $\beta,...=0,1,2,3$ are the Lorentz indices.   The coframe field is denoted as  $\vartheta^\beta$ with components (vierbeins) $\tensor{e}{_j^\beta}$. $\omega_\alpha =\ast \vartheta_\alpha$, $\tensor{\omega}{_\alpha_\beta} = \ast (\vartheta_\alpha \wedge \vartheta_\beta)$, $\tensor{\omega}{_\alpha_\beta_\gamma} = \ast (\vartheta_\alpha \wedge \vartheta_\beta \wedge \vartheta_\gamma)$,  $e:= det (\tensor{e}{_j^\beta}) = \sqrt{-g}$.  Parentheses around the indices denote antisymmetrization $[ij]:=\frac{1}{2}(ij-ji)$. The (Fock-Ivanenko) covariant exterior derivative components for spinors are $D_\alpha=\tensor{e}{^i_\alpha}D_i$ with $D_i=\partial_i+\frac{i}{4}\tensor{\Gamma}{_i^\alpha^\beta}\tensor{\Sigma}{_\alpha_\beta}$, where $\tensor{\Gamma}{_i^\alpha^\beta}$ is the Lorentz (spin) connection and $\tensor{\Sigma}{_\alpha_\beta}$ are the representations of the Lorentz generators. The metric field will be denoted as $g_{ij}(x)$ and the Minkowski metric as $\eta_{ij}$.
\cleardoublepage

\section{Introduction}
The Einstein-Cartan (EC) theory is a gauge theory of gravity, obtained by gauging the Poincaré group. The Poincaré group is a 10 parameter non-Abelian and non-compact Lie group which is the semi-direct product of 4 translations $T_4$ and 6 Lorentz transformations $SO(1,3)$, i.e., $ P(1,3) = T_4\rtimes SO(1,3)$. This approach is built over the concept of Wigner classification, where quantum particles are indexed by its invariants mass and spin which are linked to the set of translations and Lorentz transformations respectively. The gauge connection one-forms (potentials) are the spin connection $\tensor{\Gamma}{^\alpha^\beta}(x)$ linked to the Lorentz group and the orthonormal coframe field (tetrads) $\tensor{\vartheta}{^\alpha}(x)$ linked to the translations.  The translational field strength two-form $T^\alpha(x)$ is called \textit{torsion} while the rotational field strength two-form $\tensor{R}{^\alpha^\beta}(x)$ is called \textit{Lorentz} curvature. Thus the underlying arena of the Einstein-Cartan theory is called a Riemann-Cartan (RC) spacetime. The matter current three-forms associated with these gauge potentials are the canonical energy-momentum current $\tensor{\mathfrak{T}}{_\alpha}(x)$, which are the currents coupled to $\vartheta^\alpha (x)$ and sources the underlying spacetime curvature while the canonical spin current $\tensor{\mathfrak{S}}{_\alpha_\beta}(x)$ are currents coupled to the spin (Lorentz) connection and are sources of spacetime torsion \cite{doi:10.1142/p781}. In this paper, we concentrate specifically on the canonical spin current for  spin-$\frac{1}{2}$ spinorial fields which are surprisingly \textit{dual} to the chiral (axial) vector currents $\tensor{J}{_5^i}(x)$. 
\newline


In spinor electrodynamics, quantum anomalous  non-conservation of the chiral (axial) vector current even for massless fermions leads to a topological term. This was firstly explained in the context of PCAC problem of a neutral meson decaying into two photons $\pi^0 -> 2\gamma $. This is the celebrated Adler-Bell--Jackiw (ABJ) anomaly or chiral anomaly problem \cite{10.2307/j.ctt13x1c2c}.\footnote{Definitely, there exists chiral anomalies for non-Abelian gauge fields too.} In particle physics, the intrinsic angular momentum or spin plays a pivotal role. In \cite{Hayashi:1974zp} it was already claimed that the conservation law for spin current of hadrons would satisfy a PCAC (partially conserved axial vector current) like equation. It was also shown that the tensorial part of the spin current could  source \textit{massive spin-$2$ mesons}. Hence, we take this study further to set up an analogous PCAC and chiral anomaly like expression for the spin tensor.
\newline

Recently, a major breakthrough happened in condensed matter systems, where the chiral anomaly analogue was experimentally confirmed in new topological phases of matter, namely the \textit{Weyl-semimetal}. Electrons in these material behave like a Weyl fermion
which are massless relativistic particles.
A Weyl semimetal carries non-trivial topological propeties
and the Weyl fermions with opposite chiralities are
separated in momentum space and host a monopole
and an antimonopole of Berry flux in momentum space
respectively. In this situation, parallel magnetic and
electric fields ($\vec{E}\cdot \vec{B}$) can pump electrons between Weyl cones of opposite
chirality that are separated in momentum space. This
process violates the chiral charge conservation and
the number of particles of left and right chirality
are not separately conserved. This arises because of the topological term ($\frac{1}{2}  \tensor{\epsilon}{_i_j_k_l} \tensor{F}{^i^j}(x)\tensor{F}{^k^l}(x) \approx \vec{E}\cdot \vec{B}$)\cite{Zhang2016-fk}.
\newline

Recently, there have been attempts in the direction of mixed gravitational-axial anomalies in Weyl-semimetals \cite{Gooth2017-ar}. But what is important to note is that mixed gravitational-axial anomalies are purely based on General relativity (GR). GR is built over the concept of a symmetric metric field $g_{ij}(x)$ whose second-derivatives lead to the Riemann curvature tensor $R_{ijkl}(x)$. This doesn't offer a way to rather probe the interaction of spin of chiral particles with gravitational fields, which is only possible in the context of EC gravity. 
\newline

Ideally, there have not been enough concrete experimental proofs of spin current arising from a gauge perspective. Although, one should not forget that there have been some experimental signatures related to the anomalous phases of $^{3}He$ in the A phase at low temperatures having a net macroscopic spin. This also points strongly towards the existence of an asymmetric energy-momentum tensor. Also, the EC length scale
$l_{EC} \approx (\lambda_{Compton} l_{Planck}^2)^{\frac{1}{3}} \approx 10^{-29} m$ 
is seven orders of magnitude larger than the reduced Planck
scale $l_{Planck} \approx 10^{-36} m$.
PLANCK data indicate that General Relativity can be verified to scales of $\approx 10^{-28} m$ \cite{2016}. Thus, signatures of EC gravity may well be detected in the near future.
\newline 

Since, we have some hopes of spacetime being RC and also have access to such  topological materials to look into anomalies associated with chiral Weyl fermions experimentally, we take up an interesting problem which deals with finding \textit{quantum non-conservation laws for the canonical spin current of fermions}.
\subsection{Motivation}
The question that we want to address in this paper is if the conservation of spin angular momentum holds in the quantum level. If not,  then the \enquote{anomalies} arising in QFT, would lead to the breakdown of Lorentz symmetry in the quantum regime. This would imply that the \textit{algbra of Lorentz suffers from anomalies due to quantum effects}.
\newline

It is quite surprising that the canonical spin current (sources of torsion) of spinorial fields are related to the chiral (axial) currents. Hence, it is motivating enough to look for gravitational analogues of the chiral anomaly. This phenomena is not merely calculating the chiral anomaly in a curved background \citep{Kimura:1970iv, doi:10.1142/0131}, but rather a completely different anomaly where we compute the divergence of the spin current in a background Riemann-Cartan (RC) spacetime (curved + contorted) or interacting with a combination of external Abelian $U(1)$ gauge fields in a RC background.\footnote{It would be further clarified why non-Abelian fields were not mentioned here.} 
\newline 

Infact, such an anomaly has already been calculated in \citep{PhysRevLett.53.21, PhysRevLett.53.2219} and is also famously known as the \enquote{Lorentz-Anomaly}.  Thus computing these anomalies could always be useful since experimental  tests of these anomalies could be possible using the topologically non-trivial \enquote{Weyl semimetal}.
\newline

It was shown in \cite{PhysRevLett.53.21} that, purely gravitational anomalies do not exist
in $4n$ dimensions as a consequence of  charge conjugation properties of the gravitational interactions. On the contrary they are  present only in $4n+2$ dimensions \citep{kiefer2012quantum, PhysRevLett.53.21, Bertlmann2000-nx}. Infact, the existence of such chiral Lorentz anomalies leads to a breakdown of Lorentz symmetry in 4n+ 2 dimensions due to quantum effects.
\newline

Thus our main motivation is to calculate using the results obtained in \cite{PhysRevLett.53.2219}, the quantum corrections to the \textit{algebra of Lorentz} present for chiral particles interacting with a combined gravitational and $U(1)$ gauge field. Thus, by studying chiral gauge theories in curved spacetime, one can understand the interactions between \textit{intrinsic angular momentum of fermions coupled to the underlying geometry of spacetime}. 

\section{Plan of the Paper}
We firstly describe the canonical spin current of Dirac fields and its irreducible decomposition into different components. It turns out to be that only the totally \textit{antisymmetric axial vector} part survives, i.e., $^{AX}\tensor{\mathfrak{S}}{_\alpha_\beta^i}(x)=\frac{i\hbar c} {4}e\tensor{e}{^i^\gamma}\tensor{\epsilon}{_\alpha_\beta_\gamma_\delta}\tensor{J}{_5^\delta}(x).$ Then in subsection (\ref{scla}), we revisit the ETCR/ spin current algebra.
Later, we briefly go through the similarities between the ABJ anomaly and the Lorentz anomaly by considering, (i) a purely gravitational (RC) background, (ii) spin of chiral Weyl fermions interacting with a combination of an external $U(1)$ gauge field in a  background RC spacetime.
\newline 

Our main work can be found in section (\ref{MainSec}), wherein, we compute the quantum modifications to the spin current algebra which shows up as \enquote{anomalies} due to U(1) gauge fields and  Riemann-Cartan spacetime combined.
Furthermore, using these results, we reduce these spin current commutators to the Lorentz algebra for chiral Weyl fermions. This leads to a very interesting consequence that, one obtains different quantum correction terms either proportional to $\alpha$ (\enquote{Fine-structure constant}) or rather a  \textit{Dirac quantized product of the electric and magnetic charges}, if we include magnetic monopoles in Maxwell's theory. A very interesting result that we show here is that, merely swapping the Lorentz generators in the commutation relation not only leads to a sign change but rather a completely different correction terms. Thus, we show that \enquote{quantum anomalies} leads to the breakdown of local Lorentz symmetry and ultimately, alterations to the Lorentz algebra.

\section{The canonical spin current of a Spinorial field.}
\subsection{Spin currents irreducible decomposition}
The irreducible decomposition of the spin tensor ($6 \cross 4 = 24$ components) consists of a tensor part, a vector part and an axial vector part \cite{Hehl:2014eja},
\begin{equation}\label{paper02_eq:30}
\tensor{\mathfrak{S}}{_i_j^k} =^{TEN}\tensor{\mathfrak{S}}{_i_j^k} + ^{VEC}\tensor{\mathfrak{S}}{_i_j^k} + 
^{AX}\tensor{\mathfrak{S}}{_i_j^k}.
\end{equation}
The tensor piece, $^{TEN}\tensor{\mathfrak{S}}{_i_j^k}$ consists of $16$ components. The vector piece $^{VEC}\tensor{\mathfrak{S}}{_i_j^k}:=\frac{2}{3}\tensor{\mathfrak{S}}{_[_i_|_l^l}\tensor{\delta}{^k_|_j_]}$ contains 4 components and the axial vector piece being completely antisymmetric   $^{AX}\tensor{\mathfrak{S}}{_i_j_k}=\tensor{\mathfrak{S}}{_[_i_j_k_]}$ also contains 4 components.
\newline 
The Dirac Lagrangian in a Riemann-Cartan spacetime reads,
\begin{align}\label{paper02_eq:31}
\mathfrak{L}_D=\frac{\hbar c}{2}e\big(i(\bar{\psi}\gamma^\alpha D_\alpha\psi -D_\alpha\bar{\psi}\gamma^\alpha \psi)-2\frac{mc}{\hbar}\bar{\psi}\psi\big) .\end{align}
Where $\gamma^\alpha$ are the constant Dirac matrices and are the components of the Clifford-algebra valued exterior forms, i.e., 
\begin{equation*}
\gamma := \gamma_\alpha \vartheta^\alpha,       \end{equation*}
and the Fock-Ivanenko type covariant derivative acts on the spinors,
\begin{equation*}
D_i \psi(x)=   \big(\partial_i+\frac{i}{4}\tensor{\Gamma}{_i^\alpha^\beta}[\gamma_\alpha, \gamma_\beta]\big)\psi(x)
\end{equation*}
Where, $\tensor{\Sigma}{_\alpha_\beta} = \frac{i}{4}[\gamma_\alpha, \gamma_\beta]$ is the spinor representation of the Lorentz generators. 
The Dirac equation for the spinor field $\psi$ and the conjugate field $\bar{\psi}$ are,
\begin{subequations}
\begin{equation}\label{paper02_eq:32a}
i\tensor{e}{^k_\alpha}\gamma^\alpha \partial_k\psi=m\psi - i \tensor{e}{^k_\alpha} \gamma^\alpha\Gamma_k \psi ,
\end{equation}
\begin{equation}\label{paper02_eq:32b}
-i\partial_k\bar{\psi}\gamma^\alpha\tensor{e}{^k_\alpha}=\bar{\psi}m-i\bar{\psi}\Gamma_k \gamma^\alpha\tensor{e}{^k_\alpha} ,
\end{equation}
\end{subequations}
where $D_\alpha\psi=\tensor{e}{^k_\alpha}(\partial_k+\Gamma_k)\psi$ , $D_\alpha\bar{\psi}=\bar{\psi}(\partial_k-\Gamma_k)\tensor{e}{^k_\alpha}$.
\subsection{Relation between spinorial spin current and chiral (axial) vector currents}
The canonical spin current of the Dirac field are totally antisymmetric (axial part with $4$-components) and is \textit{dual to the axial vector currents}, \begin{align}\label{paper02_eq:33}
^{AX}\tensor{\mathfrak{S}}{_\alpha_\beta^i}(x) :=-2\frac{\delta\mathcal{L_D}}{\delta\tensor{\Gamma}{_i^\alpha^\beta}}=\frac{i\hbar c} {4}e\tensor{e}{^i^\gamma}\tensor{\epsilon}{_\alpha_\beta_\gamma_\delta}\bar{\psi}(x)\gamma_5\gamma^\delta\psi(x) = \frac{i\hbar c} {4}e\tensor{e}{^i^\gamma}\tensor{\epsilon}{_\alpha_\beta_\gamma_\delta} \tensor{J}{_5^\delta}(x).
\end{align}
The above Eq.\ (\ref{paper02_eq:33}), obeys the \textit{classical} conservation equation for spin angular momentum ,i.e., \cite{Hehl:2014eja}, 
\begin{equation}\label{Noether}
D_i\tensor{\mathfrak{S}}{_\alpha_\beta^i}(x) - 2\tensor{\mathfrak{T}}{_{[\alpha\beta]}} = 0.  
\end{equation}

Where $\tensor{\mathfrak{T}}{_{[\alpha\beta]}}$ is the antisymmetric part of the canonical energy-momentum current. Our goal is to study whether the above Eq. (\ref{Noether}) is subjected to \enquote{quantum non-conservation} due to anomalies. Hence, we derive an analogue of the chiral anomaly. From this point on, we work with chiral Weyl fermions which are massless and carry a handedness (chirality). 
\subsection{Spin current algebra}\label{scla}
The ETCRs for the spin current can be derived using the Schwinger quantum action principle by varying the matrix elements of Eq. (\ref{Noether}), w.r.t to the spin (Lorentz) connection $\tensor{\Gamma}{_i^\alpha^\beta}(x)$. Thus, the canonical spin current Lie algebra reads,\footnote{The spin current Lie-algebra corresponds to $\mathfrak{so}(1,3)\otimes \mathfrak{so}(1,3)$.}
\begin{align}\label{paper01_eq:17}
[\tensor{\mathfrak S}{_\alpha_\beta^0}(x),\tensor{\mathfrak S}{_\gamma_\delta^j}(x')]_{x_0=x'_0} &= 2i \big(\tensor{\eta}{_\gamma_[_\alpha}\tensor{\mathfrak S}{_\beta_]_\delta^j}(x)-\tensor{\eta}{_\delta_[_\alpha}\tensor{\mathfrak S}{_\beta_]_\gamma^j}(x)\big)\delta^3(x-x')\nonumber \\ 
&+i\bigg(\partial_i\frac{\delta\tensor{\mathfrak {S}}{_\alpha_\beta^i}(x)}{\delta\tensor{\Gamma}{_j^\gamma^\delta}(x')}-2\frac{\delta\mathfrak 
{T}_{[\alpha\beta]}(x)}{\delta\tensor{\Gamma}{_j^\gamma^\delta}(x')}\bigg).
\end{align}
The second bracket contains the Schwinger terms. The first term in the second bracket of Eq. (\ref{paper01_eq:17}) is always zero for fermions since $\frac{\delta\tensor{\mathfrak {S}}{_\alpha_\beta^i}(x)}{\delta\tensor{\Gamma}{_j^\gamma^\delta}(x')} = 0$, i.e the current is independent of the connection. For the sake of ease, one could avoid the other term in the second bracket.

\subsection{Anomalous divergence of spin currents}
Using the Lagrangian for chiral particles (chiral Lagrangian) and local Lorentz invariance, we find that the chiral Weyl spinors also satisfy the classical conservation law,   
\begin{align}\label{paper02_eq:37}
D_i\tensor{\mathfrak S}{_\alpha _\beta^i}(x)_\chi-  2\tensor{\mathfrak{T}}{_{[\alpha \beta]}}(x)_\chi = 0,
\end{align}
where subscript $\chi$ denotes chiral Weyl fermions. The spin tensor and the antisymmetric canonical energy-momentum tensors are, 
\begin{subequations}
\begin{align}\label{paper02_eq:38a}
\tensor{\mathfrak{S}}{_\alpha_\beta^i}(x)&=\frac{i e \hbar c}{4}\tensor{e}{^i^\gamma}\tensor{\epsilon}{_\alpha_\beta_\gamma_\delta}\bar{\chi}(x)\gamma_5(x)\gamma^\delta(x)\chi(x) = \frac{ i \hbar  c }{2} \tensor{\epsilon}{_\alpha_\beta_\gamma_\delta}e\tensor{e}{^i^\gamma}\tensor{J}{_5^\delta}(x),
\end{align}
\begin{align}\label{paper02_eq:38b}
\tensor{\mathfrak{T}}{_{[\alpha\beta]}}(x)&= \frac{i \hbar c}{2} e\big(\bar{\chi}\gamma_{[\alpha}D_{\beta]}\chi -D_{[\beta}\bar{\chi}\gamma_{\alpha]}\chi\big).
\end{align}
\end{subequations}
These two currents Eqs. (\ref{paper02_eq:38a} , \ref{paper02_eq:38b}) are related to \textit{Lorentz invariance} and \textit{diffeomorphism invariance} respectively. Hence, fermions interacting with purely gravitational field should possess both \textit{Lorentz invariance} and \textit{diffeomorphism invariance}. 
\newline 

In order to study the anomalies associated with the divergence of spin currents, we  concentrate on the local Lorentz symmetry. In order to achieve this we construct a gauge invariant axial vector current interacting with a \textit{background} Riemann-Cartan spacetime,
\begin{equation}\label{ps}
\tensor{J}{_5^l}(x|\epsilon)=\lim
_{\epsilon\rightarrow 0}\frac{i}{2}\bar{\psi}(x+\frac{\epsilon}{2})\gamma^5\gamma^l\exp{i\lambda\int_{x-\frac{\epsilon}{2}}^{x+\frac{\epsilon}{2}}dy^j\tensor{\Gamma}{_j^\alpha^\beta}(y)\tensor{\Sigma}{_\alpha_\beta}}\psi(x-\frac{\epsilon}{2}).
\end{equation}

Here, the Fujikawa point splitting on the fields in terms of the \textit{spin connection} has been used since $\tensor{J}{_5^\delta}(x)$ is singular. Here, $\lambda$ corresponds to a coupling associated with gravity. Substituting Eqs.\ (\ref{ps}, \ref{paper02_eq:38a}) into Eq.\ (\ref{paper02_eq:37}), we obtain the anomalous divergence of the spin tensor,\footnote{We avoid the torsion tensor coupling to the chiral currents in Eq. (\ref{paper02_eq:39}). This term potentially leads to anomalies and would be investigated more in the future.} 
\begin{equation}\label{paper02_eq:39}
D_i\tensor{\mathfrak{S}}{_\alpha_\beta^i}(x) - 2\tensor{\mathfrak{T}}{_{[\alpha\beta]}} \approx \big({\tensor{\widetilde{R}}{_\alpha^\gamma^i^j}}\tensor{R}{_\beta_\gamma_i_j}- \tensor{\widetilde{R}}{_\beta^\gamma^i^j}\tensor{R}{_\alpha_\gamma_i_j}\big)=0.
\end{equation} 
Where $\tensor{\widetilde{R}}{_\alpha_\beta^i^j}(x) = \frac{1}{2}\tensor{\epsilon}{_\alpha_\beta_\gamma_\delta}\tensor{\widetilde{R}}{^\gamma^\delta^i^j}(x)$ is the dual Lorentz curvature. The anomaly term ($\widetilde{R} R$) in the r.h.s. of Eq. (\ref{paper02_eq:39}) is always zero for $4n$ dimensions because of charge conjugation properties of the gravitational interactions. Thus the \textit{quantum corrections to the spin current algebra is non-existent in 4-dimensional spacetime}. Hence, the \textit{quantum conservation equation holds for the spin in a background RC spacetime}. This is a marked difference between the above Eq. (\ref{paper02_eq:39}) and the ABJ anomaly, where the later contains anomalous topological terms in 4-dimensional spacetime even for the massless case.

\section{Anomaly due to gravitational $+$ U(1) gauge fields and breakdown of Lorentz invariance}\label{MainSec}
In order to better understand the interaction between spin and the fundamental interactions, consider  the chiral Weyl spinors interacting with an external Abelian $U(1)$ gauge field $A_i$(x) and other non-Abelian $SU(N)$ gauge fields $\tensor{A}{^A_i}$(x) in a background Riemann-Cartan spacetime. 
\newline 
It is pretty surprising that anomalies are solely contributed by the \textit{Abelian $U(1)$ gauge fields} on a curved background. Surprisingly, the non-Abelian gauge fields don't contribute to anomalies since there is no breakdown of Poincaré invariance, This leads to the breakdown of Lorentz invariance in physical spacetime. Thus corresponding expression for the Lorentz-chiral anomaly is,  \cite{PhysRevLett.53.2219}, \footnote{The covariant derivative $;$ in Eq. (\ref{paper02_eq:41}) contains both Lorentz+affine connections. we only vary w.r.t the spin connection and don't bother about the affine part.}
\begin{align}\label{paper02_eq:41}
D_i\tensor{\mathfrak{S}}{_\alpha_\beta^i}_\chi-2\tensor{\mathfrak{T}}{_{[\alpha\beta]}}_\chi = -\frac{i q_e}{96\pi^2}\big(R\tensor{\widetilde{F}}{_\alpha_\beta}+\tensor{\widetilde{R}}{_\alpha_\beta_k_l}\tensor{F}{^k^l}+2 \tensor{F}{_\alpha_\beta_{;i}^i}\big).
\end{align}
Here, $\tensor{\widetilde{F}}{_\alpha_\beta} (x) = \frac{1}{2}  \tensor{\epsilon}{_\alpha_\beta_\gamma_\delta} \tensor{F}{^\gamma^\delta}(x)$ is the dual of the electromagnetic field strength tensor.
A major difference between Eq.\  (\ref{paper02_eq:41}) and the chiral anomaly is that the presence of a \textit{Laplacian of the electromagnetic field strength} tensor, $\tensor{F}{_\alpha_\beta_;_i^i}$. Also, the first two terms in the r.h.s of Eq.(\ref{paper02_eq:41}) shows coupling between the electromagnetic field strength and the rotational field strength (curvature) of the gravitational fields. On the contrary the chiral anomaly contains the topological term $\frac{\tensor{q}{_e^2}}{8\pi^2}\tensor{\widetilde{F}}{^i^j}\tensor{F}{_i_j}(x)$ \cite{10.2307/j.ctt13x1c2c}. Thus, the anomaly due to spin leads to several new terms which are absent in chiral anomalies.
\newline
The spin current commutators are obtained by varying Eq.\  (\ref{paper02_eq:41}) w.r.t.\ the spin connection $\tensor{\Gamma}{_j^\gamma^\delta}(x')$. We record directly the spin current commutators with anomalies included,  

\begin{align}\label{main}
[\tensor{\mathfrak S}{_\alpha_\beta^0}(x),\tensor{\mathfrak S}{_\gamma_\delta^j}(x')] &= 2i \big(\tensor{\eta}{_\gamma_[_\alpha}\tensor{\mathfrak S}{_\beta_]_\delta^j}(x)-\tensor{\eta}{_\delta_[_\alpha}\tensor{\mathfrak S}{_\beta_]_\gamma^j}(x)\big)\delta^3(x-x')\\\nonumber
&-\tensor{W}{_{\alpha\beta,\gamma\delta}^{0,j}}(x,x')\\\nonumber
&-\frac{i q_e}{24\pi^2}\big(\tensor{\widetilde{F}}{_\alpha_\beta}\tensor{e}{^j_{[\gamma}}\tensor{e}{^k_{\delta]}}(x)\partial_k\delta^3(x-x')+\frac{1}{2}\tensor{\epsilon}{_{\alpha\beta\gamma\delta}}\tensor{F}{^j^k}(x)\partial_k\delta^3(x-x')\big). 
\end{align} 
\newline 
The third term in r.h.s of Eq. (\ref{main}) is the quantum modification or the anomaly term.
By the same argument as above, the $W$ terms for spinorial fields completely vanish. Here, we explicitly write down the different components, 
\begin{align}\label{paper02_eq:43}
[\tensor{\mathfrak S}{_a_b^0}(x),\tensor{\mathfrak S}{_c_d^e}(x')]=2i\big(\tensor{\eta}{_c_[_a}\tensor{\mathfrak S}{_b_]_d^e}(x)-\tensor{\eta}{_d_[_a}\tensor{\mathfrak S}{_b_]_c^e}(x)\big)\delta^3(x-x')\\ \nonumber -\frac{i  q_e}{24\pi^2}\big(\tensor{\epsilon}{_a_b_g}E^g(x)\tensor{e}{^e_{[c}}\tensor{e}{^f_{d]}}(x)\big)\partial_f\delta^3(x-x'),
\end{align}
\begin{align}\label{paper02_eq:44}
[\tensor{\mathfrak{S}}{_a_0^0}(x),\tensor{\mathfrak {S}}{_b_0^e}(x')]&= i\tensor{\mathfrak{ S}}{_a_b^e}(x)\delta^3(x-x')\\\nonumber
&-\frac{i  q_e c}{24\pi^2}\big(B_a(x)\tensor{e}{^e_{[b}}\tensor{e}{_{0]}^f}(x)\big)\partial_f\delta^3(x-x') , 
\end{align}
\begin{align}\label{paper02_eq:45}
[\tensor{\mathfrak S}{_a_b^0}(x),\tensor{\mathfrak S}{_c_0^e}(x')]&=2i \tensor{\eta}{_c_[_a}\tensor{\mathfrak{S}}{_b_]_0^e} \delta^3(x-x')\\\nonumber
&-\frac{i q_e}{24\pi^2}\big(\tensor{\widetilde{F}}{_a_b}(x)\tensor{e}{^e_{[c}}\tensor{e}{^f_{0]}}(x)\partial_f\delta^3(x-x')-\frac{1}{2}\tensor{F}{^e^f}\partial_f\delta^3(x-x')\big),
\end{align}
\begin{align}\label{paper02_eq:46}
[\tensor{\mathfrak S}{_a_0^0}(x),\tensor{\mathfrak S}{_b_c^e}(x')]&=2i \tensor{\eta}{_a_[_b}\tensor{\mathfrak{S}}{_c_]_0^e} \delta^3(x-x')\\\nonumber
&-\frac{i  q_e c}{24\pi^2}\big(\tensor{\widetilde{F}}{_a_0}(x)\tensor{e}{^e_{[b}}\tensor{e}{^f_{c]}}(x)\partial_f\delta^3(x-x')+\frac{1}{2}\tensor{F}{^e^f}\partial_f\delta^3(x-x')\big)
\end{align}
Writing down the equations explicitly in terms of the electric and magnetic components, yields,
\begin{align}\label{paper02_eq:47}
[\tensor{\mathfrak S}{_a_0^0}(x),\tensor{\mathfrak S}{_b_c^e}(x')]&=2i \tensor{\eta}{_a_[_b}\tensor{\mathfrak{S}}{_c_]_0^e}(x) \delta^3(x-x')\\\nonumber
&-\frac{i  q_e c}{24\pi^2}\big(\frac{1}{2}(B\cross\nabla)^e\delta^3(x-x')+B_a\tensor{e}{^e_{[b}}\tensor{e}{^f_{c]}}(x)\partial_f\delta^3(x-x')\big),
\end{align}
\begin{align}\label{paper02_eq:48}
[\tensor{\mathfrak S}{_a_b^0}(x),\tensor{\mathfrak S}{_c_0^0}(x')]& = 2i \tensor{\eta}{_c_[_a}\tensor{\mathfrak{S}}{_b_]_0^0} \delta^3(x-x')\\\nonumber
&-\frac{i  q_e}{24\pi^2}\big(\frac{1}{2}(E\cdot\nabla)\delta^3(x-x')+\tensor{\epsilon}{_{abg}}E_g(x)\tensor{e}{^f_c}(x)\partial_f\delta^3(x-x')\big),
\end{align}
\begin{align}\label{paper02_eq:49}
[\tensor{\mathfrak{S}}{_a_0^0}(x),\tensor{\mathfrak {S}}{_a_0^0}(x')]&\approx	
-\frac{i q_e c}{24\pi^2}(B\cdot\nabla)\delta^3(x-x') , 
\end{align}
\begin{align}\label{paper02_eq:50}
[\tensor{\mathfrak S}{_a_b^0}(x),\tensor{\mathfrak S}{_a_b^e}(x')] \approx	 -\frac{i  q_e}{24\pi^2}(E\cross\nabla)^e\delta^3(x-x').
\end{align}
The symbol $\approx$ means there are some other terms which are not of our interest. Infact the Lorentz generators are nothing but the integral of the time component of the spin current over a space-like $3$ dimensional hypersurface, i.e.,
\begin{subequations}
\begin{align}
\tensor{\Sigma}{_b_c} := \int d^3 x \tensor{\mathfrak{S}}{_b_c ^ 0}(x) \\
\tensor{\Sigma}{_a_0} = \int d^3 x \tensor{\mathfrak{S}}{_a_0^0}(x).
\end{align}
\end{subequations}

Where $\tensor{\Sigma}{_a_b}$ are the Lorentz rotations while $\tensor{\Sigma}{_a_0}$ are the Lorentz boosts. 

Integrating Eq.\  (\ref{paper02_eq:48}) w.r.t.\ x \& x' and using the Gauss theorem yields,
\begin{equation}\label{paper02_eq:51}
[\Sigma_{ab},
\Sigma_{c0}]= \big(\tensor{\eta}{_{ac}}\Sigma_{b0}-\tensor{\eta}{_{bc}}\Sigma_{0a}\big)-\frac{i \hbar c}{12\pi} \alpha + ... .
\end{equation}
Where $\alpha = \frac{q_e^2}{\hbar c}$ is the \textit{fine structure constant}. 

This is a remarkable result, since the second term in Eq. (\ref{paper02_eq:51}) is the alteration to the Lorentz algebra due to  quantum corrections. Hence, the the existence \textit{quantum anomalies} leads to the breakdown of local Lorentz symmetry.

Infact,  integrating Eq.\ (\ref{paper02_eq:49}) w.r.t.\ x and x' and using the Gauss theorem yields a very surprising result,
\begin{equation}\label{paper02_eq:52}
[\Sigma_{0a}, \Sigma_{0a}]= -\frac{i \hbar c}{12\pi}n. 
\end{equation}
Where $n= \frac{2 q_e q_m}{\hbar c} \in \mathbb{Z} $ is the \textit{Dirac quantization}. $q_m$ is the magnetic charge. Ideally, the algebra of two same boost generators is zero. If we don't consider magnetic monopoles, this yields the standard result, which is zero.
\section{DISCUSSION AND CONCLUSION}
Firstly we infer that the purely gravitational anomaly term proportional to ${\tensor{\widetilde{R}}{_\alpha^\gamma^i^j}}\tensor{R}{_\beta_\gamma_i_j}- \tensor{\widetilde{R}}{_\beta^\gamma^i^j}\tensor{R}{_\alpha_\gamma_i_j}$, vanishes identically in four dimensions.  This is related to the charge conjugation properties of the gravitational field. Hence a \textit{quantum conservation law holds for spin currents in four dimensions} and the spin current algebra contains no anomalous terms. 
\newline

We see that $[\mathfrak{S}_r, \mathfrak{S}_{r,b}]$ \footnote{Here, $\mathfrak{S}_{r,b}$ implies either $\mathfrak{S}_r$ or $\mathfrak{S}_b$} always contains correction terms related to the $\vec{E}$ fields. On the contrary, $[\mathfrak{S}_b, \mathfrak{S}_{r,b}]$, contains terms proportional to $\vec{B}$ fields as closure failure. Thus local active transformations performed on  Weyl fermions in the presence of \textit{$U(1)$ gauge fields}, produces, apart from a mere sign change, completely different \textit{fields} and \textit{charges} (electric or magnetic). Hence, anomalies are sensitive to the order of performing symmetry transformations. 
\newline


A very interesting result is Eq. (\ref{paper02_eq:52}), where the rotation-boost commutation contains a correction term proportional to $\frac{q_e^2}{\hbar c}$ or the \enquote{fine-structure constant} .
\newline 

The commutation between the rotation and boosts generators contains a quantum correction term proportional to the electric charge squared. While the same boost generators close on a non-trivial topological number $n \in  \mathbb{Z}$ which is the Dirac quantized product of electric and magnetic charges.
\newline

If magnetic charges are not taken into consideration, Eqs.\ (\ref{paper02_eq:47}, \ref{paper02_eq:48}) (rotation-boost or boost-rotation commutators)  contain the Maxwell \textit{source equations}.
The \textit{source free} Maxwell equations show up in  - Eqs.\ (\ref{paper02_eq:49}, \ref{paper02_eq:50}) (rotation-rotation or boost-boost commutators). This could have some important implications and would be investigated in future works.
\newline

Similar anomalous current commutation relations  were found for the vector and axial vector currents (cf. \cite{PhysRev.187.1935}). Here too, the additional anomaly terms corresponded to electric and magnetic fields or charges and the current commutation were order sensitive.
\newline 

Our further goal would be to rather come up with viable experimental approaches to look for chiral-Lorentz anomaly signatures using Weyl semimetals. This shall be investigated in the near future.

\begin{acknowledgments}
I am indebted to my supervisor Prof.\ Dr.\ Claus Kiefer who constantly guided and supported me to work on this problem accompanied by many helpful and insightful discussions. I am also indebted to Prof.\ Dr.\ Friedrich W. Hehl for for consistently guiding me with numerous insightful discussions and helpful remarks. This project was partly supported by the Bonn–Cologne Graduate School of Physics and Astronomy scholarship.
\end{acknowledgments}

\bibliography{References}

\end{document}